# Orbital angular momentum light frequency conversion and interference with quasi-phase matching crystals

Zhi-Yuan Zhou[1,2], Dong-Sheng Ding[1,2], Yun-Kun Jiang[3], Yan Li[1,2], Shuai Shi[1,2],

Xi-Shi Wang[4], Bao-Sen Shi[1,2],[*] and Guang-Can Guo[1,2]

[1]*Key Laboratory of Quantum Information, University of Science and Technology of China, Hefei, Anhui 230026, China*

[2]*Synergetic Innovation Center of Quantum Information & Quantum Physics, University of Science and Technology of China, Hefei, Anhui 230026, China*

[3]*College of Physics and Information Engineering, Fuzhou University, Fuzhou 350002, China*

[4]*State Key Lab. of Fire Science, University of Science & Technology of China, Hefei, Anhui 230026, China*

[*]*Corresponding author: drshi@ustc.edu.cn*

Light with helical phase structures, carrying quantized orbital angular momentum (OAM), has many applications in both classical and quantum optics, such as high-capacity optical communications and quantum information processing. Frequency conversion is a basic technique to expand the frequency range of fundamental light. The frequency conversion of OAM-carrying light gives rise to new physics and applications such as up-conversion detection of images and high dimensional OAM entanglements. Quasi-phase matching (QPM) nonlinear crystals are good candidates for frequency conversion, particularly for their high-valued effective nonlinear coefficients and no walk-off effect. Here we report the first experimental second-harmonic generation (SHG) of OAM light with a QPM crystal, where a UV light with OAM of $100\hbar$ is generated. OAM conservation is verified using a specially designed interferometer. With a pump beam carrying an OAM superposition of opposite sign, we observed interesting interference phenomena in the SHG light; specifically, a photonics gear-like structure is obtained that gives direct evidence of OAM conservation, which will be very useful for ultra-sensitive angular measurements. We also develop a theory to reveal the underlying physics of the phenomena. The methods and theoretical analysis shown here are also applicable to other frequency conversion processes, such as sum frequency generation and difference-frequency generation, and may also be generalized to the quantum regime for single photons.

PACS numbers: 42.65.Ky,42.50.Tx, 42.25.Hz, 42.70.Mp

Orbital angular momentum (OAM) in light is a very useful degree of freedom that has no dimensional limitation, and has been widely studied in both classical and quantum optics fields since first introduced by Allen in 1992 [1]. It has been shown [1] that a light beam with a helically phase $e^{il\theta}$ in the azimuthal direction carries $l\hbar$ units of OAM per photon. OAM light has been widely used in many fields, such as optical manipulation [2-5], optical trapping [6], optical tweezers [7], optical vortex knots [8], imaging [9], astronomy [10], free-space information transfer and communications [11], and quantum information processing [12-19].

The interaction of OAM light with matter, such as nonlinear crystals [12, 20-22] and atomic vapors [23-25], produces many interesting phenomena in contrast to those obtained using Gaussian beams. Allen [20, 21] has demonstrated the OAM transformation and conservation in frequency conversion in LBO crystal. Zeilinger's group [12] has realized high-dimensional OAM entanglement in the spontaneous parametric down-conversion processes. In all these nonlinear interaction processes, the total OAM conservation of light plays a very important role. The frequency conversion of OAM lights will be very useful in up-converting detection of images [26] and generating of OAM light from a fundamental OAM light at special wavelengths (in the UV or mid-infrared frequency domains), which are hard to produce them with traditional method. For nonlinear processes with crystals, the benefits from quasi-phase matching (QPM) when compared with birefringence phase matching make QPM crystals good candidates for frequency conversion of OAM light, particularly for their high-valued effective nonlinear coefficients and no walk-off effect. Then some important questions are coming naturally: can we use QPM crystals for nonlinear frequency conversion of OAM light? Whether the total OAM of light is conserved in such nonlinear processes? Is frequency conversion of OAM superposition state possible? So far, no experimental work has been reported on such frequency conversion processes, although one theoretical study has appeared on the process of sum frequency generation (SFG) and second-harmonic generation (SHG) [27].

In this work, the previous posed questions are answered, we report the first experimental generation of OAM-carrying UV light in SHG with a QPM type-I PPKTP crystal. We demonstrate the conservation of OAM in the SHG process, which concurs with the theory of Ref. 27. Moreover, we observe a very interesting interference phenomenon by transforming the pump light into a hyper-superposition of polarization and OAM states. We directly see a photonic gear-like structure that has never before been observed or discussed in three-wave mixing processes. This phenomenon can be regarded as direct evidence of OAM conservation. The photonic gear can be rotated by rotating the pump-beam polarization, an effect that can be used for ultra-sensitive angular measurements. These observations can be well explained by the theory we have developed. The method we demonstrate here provides a new way to generate OAM light via frequency conversion in QPM crystals. Moreover, because of its low diffraction, UV light can enhance the resolution of OAM light-based imaging. Using the SHG process in OAM light-based ultra-sensitive angular measurements [18], resolutions can be further enhanced by a factor of 2. Our approach may also be used in sum frequency generation (SFG) or difference-frequency generation (DFG) [28] at the single-photon level. This will be very useful for quantum information processing using the OAM degrees of freedom of photons.

We first demonstrate OAM conservation in the SHG process. Figure 1 shows the different blocks used in our experiments; blocks a, b, and e are used in the conservation demonstration. Block-a is used to generate OAM light with the proper polarization using vortex phase plates (VPPs, from RPC photonics). Block-b performs frequency conversion and comprises two lenses (both have the same focus length of 125mm), a type-I PPKTP crystal, and a UV filter used to remove the pump light. The 1 mm×2 mm×10 mm PPKTP crystal, supplied by Raicol Crystals, was designed for SHG of wavelengths from 795 nm to 397.5 nm. Both end faces are anti-reflection coated for these two wavelengths. The measured nonlinear conversion efficiency for the PPKTP crystal is 1%/W for a Gaussian pump mode. In our experiments, the pump power was 10 mW, which produces an SHG light power of around 1 μW. The laser light we used was from a continuous wave Ti: sapphire laser (Coherent, MBR 110, less than 100 KHz line width when locked). The measured phase matching temperature of the crystal is 64.3°C. The temperature of the crystal was controlled with a semiconductor Peltier cooler with stability of ±2 mK. Block-e is a specially designed balanced interferometer for generating light in a superposition of OAM states of the form $|l\rangle + e^{i\varphi}|-l\rangle$ given the input state $|l\rangle$. The output of this interferometer, taken using a commercial charge coupled device (CCD) camera, has an intensity distribution showing a radially symmetric pattern with $2l$ maxima arranged in a ring. Counting the numbers of maxima in the pattern yields the value of OAM for the pump and SHG beam.

As shown in Ref. 27, the OAM of the SHG light is doubled compared with that of the pump beam if the Gouy phases are the same for both. The experimental results are shown in Figure 2 paired with corresponding results from theoretical simulations. In the top set of panels in Figure 2, image-a gives the intensity profile of the pump beam with $l = 2$, image-b gives the interference pattern obtained by directing the beam into block-e, image-c gives the intensity profile for the SHG beam, and image-d gives the corresponding interference pattern. We found that the number of maxima of the SHG light is exactly twice the number of the pump beam, which verifies the conservation of OAM in the SHG process. The bottom set of panels is similar to the top set, but with a pump beam with $l = 20$. The pump beam has 40 maxima in the interference pattern; the corresponding SHG light has 80 maxima. The value of $l$ can be further increased in principle; it is limited by the dimensions of the crystal. For crystals with larger cross-sections, $l$ can be much large

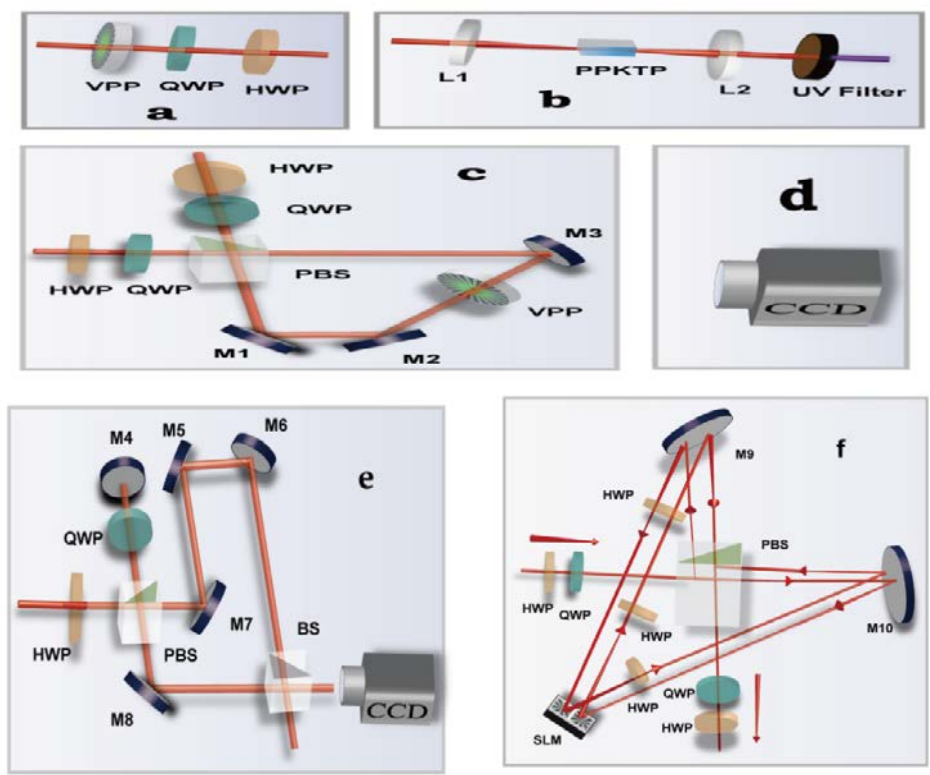

**Figure 1.** Different blocks used in experiments. HWP (QWP): half-wave plate (quarter-wave plate); PBS: polarization beam splitter; BS: beam splitter; M1-10: mirrors; L1-2: lenses; VPP: vortex phase plate; PPKTP: periodically poled KTP; SLM: spatial light modulator; CCD: charge coupled device camera. Block-a is used to prepare a single OAM pump beam with proper polarization using VPPs. Block-b is the SHG part. Block-c is a modified Sagnac interferometer for generating the superposition of OAM states of opposite sign using VPPs. Block-d is a charge coupled detector array used in direct imaging of the intensity profiles of the output of block-b. Block-e is a specially designed balanced interferometer used to determine the OAM value of the input light. Block-f has the same function as block-c, but uses an SLM instead of a VPP.

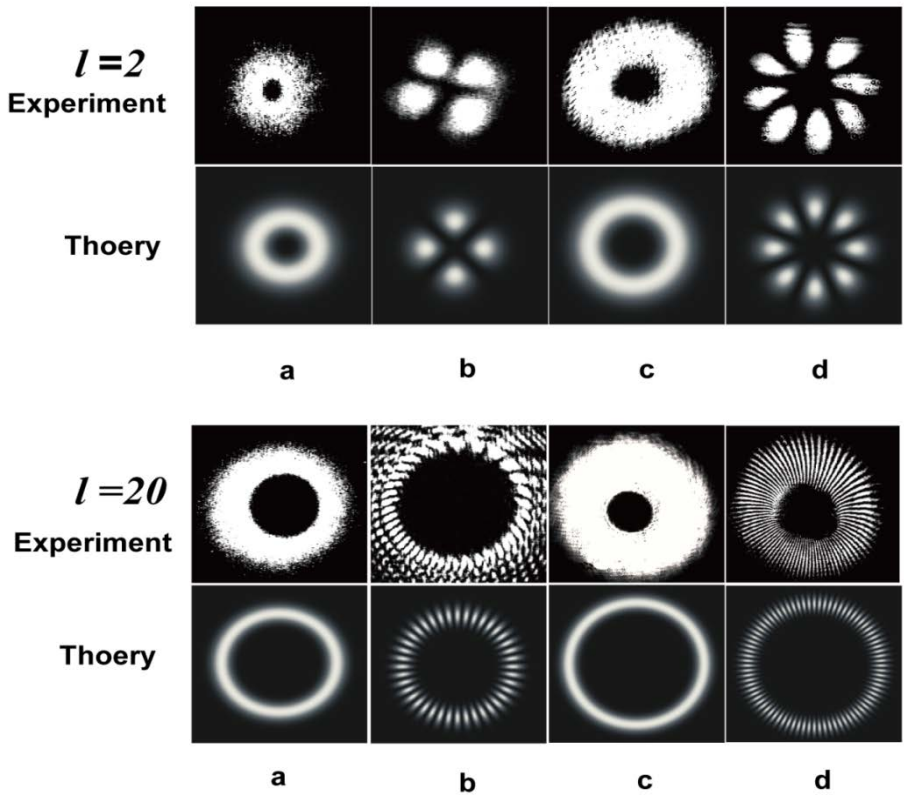

**Figure 2.** Experimental demonstration of OAM conservation in the SHG process. Column a and b give the intensity profile and interference pattern for the pump beam; Column c and d give the corresponding image for the SHG light. The top and bottom sets of panels are for pump beams carrying OAM of 2 and 20, respectively. Corresponding theoretical images are paired below the experimental images.

When we use pump-beam light with a hyper-superposition of polarization and OAM states for SHG, a photonic gear-like structure is obtained. Before showing the experimental results, we first give a detailed theoretical description. We use quantum mechanics to describe the transformation of light in block-c or -f, and a configuration similar to that presented in Refs. 17, 29, and 30 is used in our experiment. We assume that the input beam of the interferometer is in a Gaussian mode and is

polarized in the horizontal direction. The input state can be expressed as

$$|\Psi_{in}\rangle = |H\rangle \otimes |0\rangle, \qquad (1)$$

where $|H\rangle$ denotes the polarization degree of freedom and $|0\rangle$ represents the OAM degrees of freedom. After passing through block-c (or block-f), the light is transformed into the state

$$|\Psi_{out}(\theta,\delta)\rangle = \frac{1}{\sqrt{2}}[(\cos(2\delta)|H\rangle - \sin(2\delta)|V\rangle) \otimes |l\rangle - e^{-i(4\theta+\pi/2)}(\sin(2\delta)|H\rangle + \cos(2\delta)|V\rangle) \otimes |-l\rangle], \quad (2)$$

where $\theta$ and $\delta$ are the angles of the fast axis of the half-wave plate (HWP) with respect to the vertical axis at the respective input and output ports of the block, and $l$ is the OAM quantum number imprinted on the two counter-propagated beams in the interferometer. The output SHG light is in the form (see Supplementary Information for details)

$$E_{SHG} \propto \sin^2(2\delta) LG_0^{2l} - e^{-i8\theta}\cos^2(2\delta) LG_0^{-2l} + \Gamma \sin(4\delta) e^{-i(4\theta+\pi/2)} LG_l^0, \quad (3)$$

where $\Gamma$ is a constant of renormalization. This expression shows that the output of the SHG light is a superposition of OAM states of $2l$, 0, and $-2l$ that depends on the angle of $\delta$. We now focus on the case $\delta = \pi/8$; apart from a relative phase of $8\theta$, the first two terms have the same amplitude. As mentioned before, an interference pattern with $4l$ maxima is generated in the intensity distribution of the outer ring, giving direct evidence of OAM conservation in the SHG process. More interesting is that the interference pattern can be rotated if the phase $\theta$ is changed, indicating that the total phase of the pump beam is preserved in the SHG process. This behaviour is similar to a mechanical gear—when $\theta$ changes by $\pi/4$, the pattern rotates through angle $\pi/2l$—and can be exploited for ultra-sensitive measurements of angles. Furthermore, by changing $\delta$, we can switch easily between states $|2l\rangle$, $|-2l\rangle$, and $|2l\rangle + e^{i\varphi}|-2l\rangle$, thus presenting a means to perform optical switching between different OAM states (The third term in equation (3) can be removed by spatial filtering).

Based on the above theoretical analysis, we performed our experiments using blocks c, b, d and blocks f, b, d. The experimental and theoretical results from the former setup for OAM light with $l=2$ and $l=20$ are shown in Figure 3(a) and (b), respectively. The pump power is about 10 mW, and the temperature of the PPKTP is kept at the QPM point. The first and third images are the respective interference patterns of the pump light and corresponding SHG light observed by a CCD camera directly after block-b; the second and fourth images are the corresponding theoretical patterns and show good agreement with the experimental results. As the Supplementary Information shows, the generated modes of the SHG beam are not standard Laguerre-Gaussian (LG) modes at the position where they are emitted. After propagating to the far-field, they evolve into the standard LG modes. There are nine maxima in the SHG intensity profile for $l=2$, created via the interference among modes $LG_0^4$, $LG_0^{-4}$, and $LG_2^0$. For small OAM, the diffraction of the $LG_l^0$ mode is close to the $LG_0^{2l}$ mode (the first two modes have the same diffraction property), hence both will overlap partially in the far-field. For large OAM, however, these two modes have different diffraction properties, hence almost separate from each other. In addition, the intensity redistribution makes the mode $LG_l^0$ hard to observe. For $l=20$, the interference from state $|40\rangle+e^{i\beta}|-40\rangle$ can be seen very clearly, but the $LG_{20}^0$ mode is blurred and hard to see; only a dim point can be distinguished at the centre, and hence we cannot observe the multi-ring structure. By rotating the angle of the HWP in the input port of the interferometer in block-c, a rotation in the output image is observed. We also find that the image of the SHG light is clearer than the input; this is because waves of shorter wavelength are diffracted less.

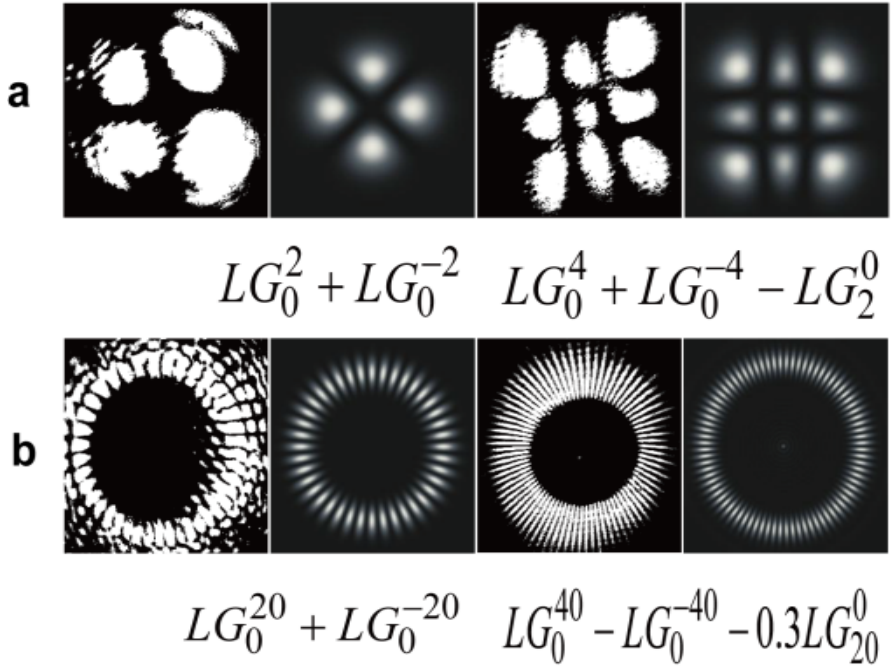

$LG_0^2+LG_0^{-2} \quad LG_0^4+LG_0^{-4}-LG_2^0$

$LG_0^{20}+LG_0^{-20} \quad LG_0^{40}-LG_0^{-40}-0.3LG_{20}^0$

**Figure 3.** SHG results with an input light in state $|l\rangle+e^{i\beta}|-l\rangle$ with (a) $l=2$ and (b) $l=20$ generated via the VPP using block-c. The first and third images of each row are the respective interference patterns for the pump light and the corresponding SHG light obtained directly from the block-b using CCD camera. The second and fourth images are the corresponding theoretical patterns.

To generate light with high OAM, or a general superposition of OAM, we also perform the experiment using a spatial light modulator (SLM, PLUTO, phase only SLMs, which has an active area of 15.36×8.64 mm$^2$, a pixel pitch size of 8 μm, and a total number pixels of 1920×1080). The setup for generating the pump state is shown in block-f. By applying this configuration, we can generate a pump light in the state $|l\rangle+e^{i\beta}|-l\rangle$ or $|l_1\rangle+e^{i\beta}|-l_2\rangle$; the first state is the same as that prepared using VPPs, the second is an asymmetrical state. Using this configuration, the two counter-propagated beams have the same optical length with an intrinsically stable phase between them. The results are shown in Figure 4. The first image in each row is the phase diagram of the SLM for generating LG modes with specific $l$ value; the other images are similarly arranged as in Figure 3. For rows a–e, the pump light is in the state $|l\rangle+e^{i\beta}|-l\rangle$ with $l=2,3,9,20,50$, respectively; row f gives images for the pump beam in the state $|l_1\rangle + e^{i\beta}|-l_2\rangle$ with $l_1 = 7$, $l_2 = 8$.

For each $l$, the number of maxima in the intensity profile of the outer ring is the same as theoretically predicted. For large $l$, there is an additional SHG light in the central region (rows e, f) arising from limitations in creating the mode at the SLM (which are arising from high-order LG mode with the same OAM and unmodulated light, respectively). There would be no such artefact if high-quality VPPs were used (see row b in Figure 3 for comparison). In row e, the OAM of UV is 100, corresponding to 200 maxima in its intensity profile. We cannot increase the OAM further because the SLM cannot operate at high powers; also, our CCD camera has a limited resolution.

For the asymmetrical state $|7\rangle + e^{i\beta}|-8\rangle$, as imaged in row f, the generated modes in the far-field are $LG_0^{14}$, $LG_0^{-16}$, and $LG_7^{-1}$; the interference pattern of the pump has 15 maxima, whereas the SHG light has 30 maxima. The pattern is not sufficiently clear as the LG modes with different absolute values of $l$ have different diffraction properties. Hence the two modes do not completely overlap in the far-field.

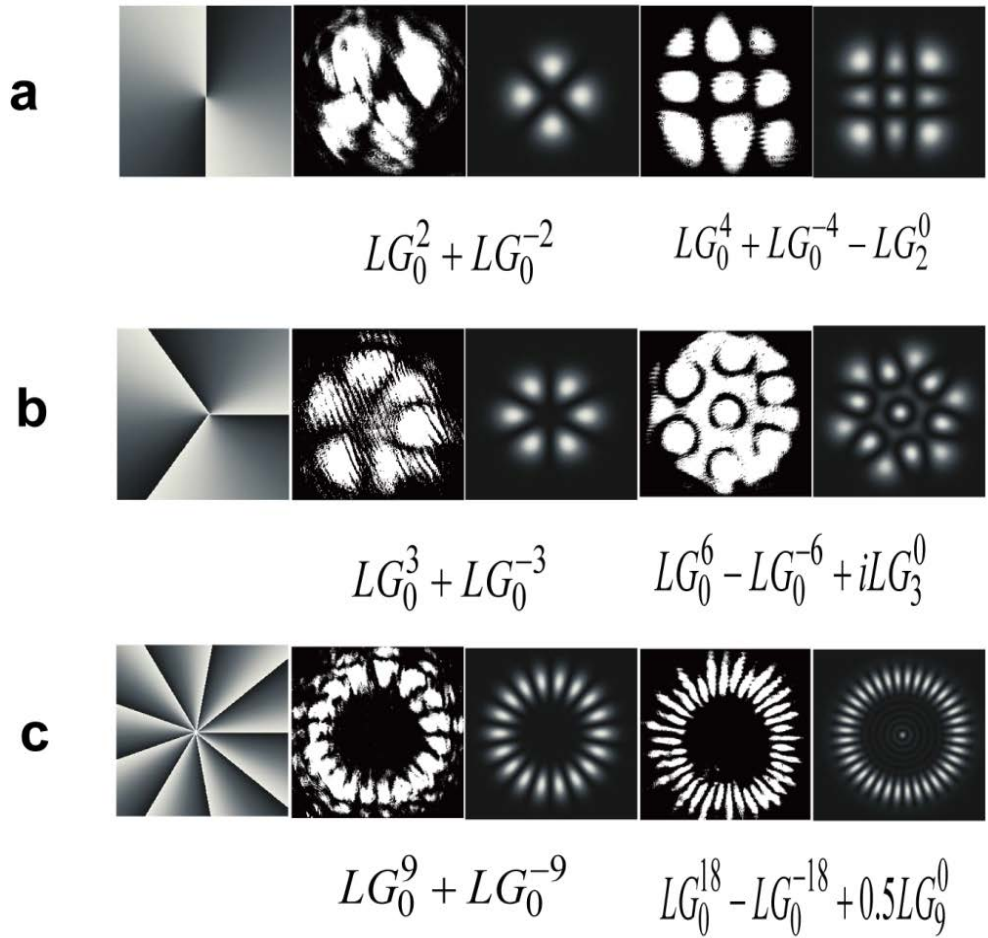

a  $LG_0^2 + LG_0^{-2}$   $LG_0^4 + LG_0^{-4} - LG_2^0$

b  $LG_0^3 + LG_0^{-3}$   $LG_0^6 - LG_0^{-6} + iLG_3^0$

c  $LG_0^9 + LG_0^{-9}$   $LG_0^{18} - LG_0^{-18} + 0.5LG_9^0$

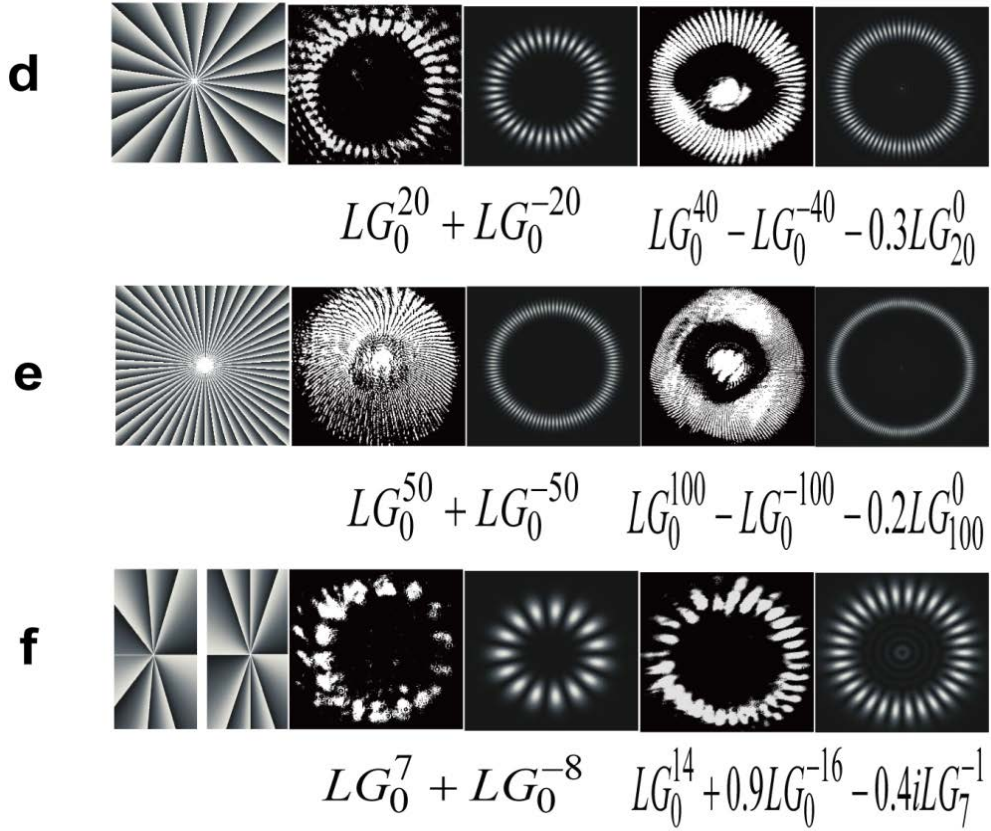

**Figure 4.** SHG results with pump states of the form $|l\rangle + e^{i\beta}|-l\rangle$ and $|l_1\rangle + e^{i\beta}|-l_2\rangle$ generated using block-f via an SLM. Rows a–e present images corresponding to states $|l\rangle + e^{i\beta}|-l\rangle$ with $l = 2, 3, 9, 20, 50$ for the pump beam; row f corresponds to state $|l_1\rangle + e^{i\beta}|-l_2\rangle$ with $l_1 = 7$, $l_2 = 8$.

The first image in each row is the phase diagram of SLM for generating a specific OAM-carrying light. The second and fourth images are the respective interference patterns for the pump light, projected onto the diagonal polarization direction, and the SHG light, directly observed after block-b using CCD camera. The third and fifth images are the corresponding theoretical patterns.

In summary, two experiments using the type-I QPM PPKTP crystal have been conducted to investigate OAM transformation and conservation in the SHG process. In the first of the two, we verified that OAM is conserved in the SHG by directing the pump and SHG OAM light into a specially designed balanced interferometer. The conservation law is confirmed by counting the maxima in the interference intensity profile. As the QPM crystal has a high-valued effective nonlinear coefficient and no walk-off effect, it provides a new method to generate OAM light by frequency conversion in QPM crystals. The image resolution depends on the wavelength of light used; shorter wavelengths yield better image resolutions. UV OAM light would be suitable for OAM light-based phase imaging. In the second of the experiments, we observed a very interesting interference phenomenon when pumping the PPKTP crystal with a superposition of two OAM states of opposite sign. The output SHG light intensity profile depended on the polarization of the pump light. A photonics gear-like structure is observed that can be rotated when the pump polarization is rotated. This effect can be used for remote sensing, OAM light-based ultra-sensitive angular measurements, and detection of spinning objects [31]. This interference effect can also be used for optical switching between different SHG patterns generated by controlling the polarization of the pump beam. We also gave analytical expressions for propagation of the SHG light for the tight focus approximation. All experimental phenomena can be well explained within the theory we have developed. For SFG and DFG conversions, the method is not limited to just the classical regime, and can be extended into the quantum regime for single photons.

**Acknowledgements**

This work was supported by the National Fundamental Research Program of China (Grant No. 2011CBA00200), the National Natural Science Foundation of China (Grant Nos. 11174271, 61275115, 10874171), and the Innovation Fund from the Chinese Academy of Science.

**Supplementary Information**

**Theoretical analysis of the generated SHG light**

Laguerre-Gaussian (LG) modes are characterized by two indices, the azimuthal index $l$ and the radial index $p$, where $l$ is the number of $2\pi$ cycles in phase around the circumference and $p$ is related to the number of radial nodes [1]. The normalized wave function of an LG mode in cylindrical coordinates is given by

$$LG_p^l(r,\varphi,x) = \sqrt{\frac{2p!}{\pi(|l|+p)!}} \frac{1}{w(x)} \left(\frac{\sqrt{2}r}{w(x)}\right)^{|l|} L_p^{|l|}\left(\frac{2r^2}{w(x)^2}\right)$$
$$\exp\left(-\frac{r^2}{w(x)^2}\right) \exp\left(ik\frac{r^2}{2R(x)}\right) \exp\left(-i[2p+|l|+1]\varsigma(x)\right) \exp(-il\varphi) \quad (1)$$

where $w(x)$ is the beam radius at position $x$, which is the axial distance from the beam waist; $L_p^{|l|}$ is the generalized Laguerre polynomials; $k = 2\pi/\lambda$ the wave number; $R(x)$ the radius of curvature of the wavefronts for the beam; $\varphi$ denotes the azimuthal angle; and $\varsigma(x)$ the Gouy phase, which adds an extra contribution to the phase.

To investigate the frequency conversion processes in QPM crystals, coupled wave functions are used to describe the interaction of these waves [2, 3]

$$\begin{cases} \dfrac{dE_{1z}}{dx} = iK_1 E_{2z}^* E_{3z} e^{-i\Delta kx} \\ \dfrac{dE_{2z}}{dx} = iK_2 E_{1z}^* E_{3z} e^{-i\Delta kx} \\ \dfrac{dE_{3z}}{dx} = igK_3 E_{1z} E_{2z} e^{i\Delta kx} \end{cases} \quad (2)$$

where $E_{jz} = LG_p^l(r,\varphi,x)\,(j=1,2,3)$ represent the amplitudes of the three waves involved and $z$ denotes the polarization; the $K$-coefficients are $K_j = \dfrac{2\omega_j d_{33}}{\pi n_j c}(j=1,2,3)$ with $\omega$ the angular frequency, $n$ the refractive index, $c$ the speed of light, $d_{33}$ the nonlinear coefficient of KTP; the values of the degeneracy factor are $g = 0.5$ for SHG, as frequency is two-fold degenerate and $g = 1$ for SFG and DFG, and $\Delta k = k_{3z} - k_{1z} - k_{2z} - G_m$, where $k$ represents the wavenumber vector. For phase matching, $\Delta k = 0$, which means the momentum mismatch is fully compensated by the reciprocal vector $G_m$ of the QPM crystals.

We use the language of quantum mechanics to describe the transformation of light via blocks c or f. We use a configuration similar to that described in Refs. 17, 29, and 30. We assume a Gaussian spatial mode polarized in the horizontal direction for the input beam of the interferometer. The input state can be expressed in the tensor product form

$$|\Psi_{in}\rangle = |H\rangle \otimes |0\rangle, \quad (3)$$

in which the first state, here $|H\rangle$, gives the polarization degrees of freedom and the second, here $|0\rangle$, gives the OAM degrees of freedom. The function of the half- and quarter-wave plates is to apply a unitary rotation to the polarization degrees of freedom. We use the Jones calculus notation, with convention

$$|H\rangle = \begin{pmatrix} 1 \\ 0 \end{pmatrix}, \quad |V\rangle = \begin{pmatrix} 0 \\ 1 \end{pmatrix}. \quad (4)$$

The functions of the quarter- and half-wave plates, whose fast axes are at angles $\varphi$ and $\theta$ with respect to the vertical axis, are given by the respective 2×2 matrices

$$\hat{U}_{QWP}(\varphi) = \dfrac{1}{\sqrt{2}} \begin{pmatrix} i - \cos(2\varphi) & \sin(2\varphi) \\ \sin(2\varphi) & i + \cos(2\varphi) \end{pmatrix},$$
$$\hat{U}_{HWP}(\theta) = \dfrac{1}{\sqrt{2}} \begin{pmatrix} \cos(2\theta) & -\sin(2\theta) \\ -\sin(2\theta) & -\cos(2\theta) \end{pmatrix}, \quad (5)$$

After passing through the plates, the polarization of the beam becomes

$$\hat{U}_{QWP}(\varphi) \bullet \hat{U}_{HWP}(\theta) \begin{pmatrix} 1 \\ 0 \end{pmatrix} = a(\theta,\varphi)|H\rangle + b(\theta,\varphi)|V\rangle, \quad (6)$$

where

$$a(\theta,\varphi) = \frac{1}{\sqrt{2}}[i\cos(2\theta) - \cos(2\varphi - 2\theta)],$$
$$b(\theta,\varphi) = \frac{1}{\sqrt{2}}[-i\sin(2\theta) + \sin(2\varphi - 2\theta)] \quad (7)$$

After propagating through the interferometer in the opposite direction, the output state of the light will be

$$|\Psi_{out}(\theta,\varphi)\rangle = a(\theta,\varphi)|H\rangle \otimes |l\rangle + b(\theta,\varphi)|V\rangle \otimes |-l\rangle. \quad (8)$$

This is a hyper-superposition of the polarization and the OAM degrees of freedom, where $l$ is the quantum number for OAM imprinted on the beam in the different propagation directions. We fix $\varphi = \pi/4$ to distribute equally the intensity of the input beam. If we neglect the common phase $e^{i(2\theta + \pi/2)}$, Eq. (8) becomes

$$|\Psi_{out}(\theta)\rangle = \frac{1}{\sqrt{2}}(|H\rangle \otimes |l\rangle + e^{-i(4\theta + \pi/2)}|V\rangle \otimes |-l\rangle). \quad (9)$$

For light passing through a half-wave plate whose fast axis is at an angle $\delta$ with respect to the vertical axis before pumping the QPM crystal, the states of (9) are transformed to

$$|\Psi_{out}(\theta,\delta)\rangle = \frac{1}{\sqrt{2}}[(\cos(2\delta)|H\rangle - \sin(2\delta)|V\rangle) \otimes |l\rangle - e^{-i(4\theta + \pi/2)}(\sin(2\delta)|H\rangle + \cos(2\delta)|V\rangle) \otimes |-l\rangle]. \quad (10)$$

In type-I QPM crystals, only the light polarized in the vertical direction is involved in the SHG process, so the projection of state (10) in the vertical direction is

$$|\Psi_{out}(\theta,\delta)\rangle_V = -\frac{1}{\sqrt{2}}(\sin(2\delta)|l\rangle + e^{-i(4\theta + \pi/2)}\cos(2\delta)|-l\rangle). \quad (11)$$

Considering the coupled wave functions in Eq. (2), there will be three pumping conditions arising from the form of equation (11), namely the interactions: (a) $|l\rangle$ with $|l\rangle$, (b) $|-l\rangle$ with $|-l\rangle$, and (c) $|l\rangle$ with $|-l\rangle$. Therefore, the generated SHG light will be in the form

$$E_{SHG} \propto \sin^2(2\delta)LG_0^{2l} - e^{-i8\theta}\cos^2(2\delta)LG_0^{-2l} + \Gamma\sin(4\delta)e^{-i(4\theta + \pi/2)}LG_l^0 \quad (12)$$

where $\Gamma$ is a normalization constant.

For type-II QPM crystals, the generated SHG light has the form

$$E_{SHG} \propto \sin(4\delta)LG_0^{2l} - e^{-i8\theta}\sin(4\delta)LG_0^{-2l} + \Gamma\cos(4\delta)e^{-i(4\theta + \pi/2)}LG_l^0. \quad (13)$$

Equations (12) and (13) are phenomenological results valid in the far-field approximation. To be more precise, we derive an analytical expression in an approximation valid for tight focusing or short crystal lengths. This means that the crystal length is much smaller than the Rayleigh range of the pump beam or the interaction length is very small because of the tight focusing. At these approximations, the SHG light generated for the three cases for the given spot size can be expressed as

$$E_{SHG}(r_0,\varphi_0,0) \propto \begin{cases} LG_0^l(r_0,\varphi_0,0) * LG_0^l(r_0,\varphi_0,0) & \text{case a} \\ LG_0^{-l}(r_0,\varphi_0,0) * LG_0^{-l}(r_0,\varphi_0,0) & \text{case b} \\ LG_0^l(r_0,\varphi_0,0) * LG_0^{-l}(r_0,\varphi_0,0) & \text{case c} \end{cases} \quad (14)$$

Cases a and b are the same, including the sign of $l$. Within the paraxial approximation, the propagation of the generated SHG light through a stigmatic ABCD optical system can be analysed aided by the Collins integral [4]

$$E(r,\alpha,z) = \frac{i}{\lambda B}\exp(-ikz)\int_0^{2\pi}\int_0^{\infty} E_{SHG}(r_0,\alpha_0,0)$$
$$\exp\left\{-\frac{ik}{2B}\times[Ar_0^2 - 2rr_0\cos(\alpha-\alpha_0) + Dr^2]\right\}r_0 dr_0 d\alpha_0 \quad (15)$$

Combined with equation (1) and substituting equation (14) into (15), the field at distant $z$ from the source point is of the form

$$E_{SHG}(r,\alpha,z) = \frac{2}{|l|!}\frac{1}{w_0^2}\left(\frac{\sqrt{2}}{w_0}\right)^{2l}\frac{i}{\lambda B}\exp(-\frac{ikD}{2B}r^2 - ikz)\exp(-\frac{k^2 r^2}{4\xi B^2})$$

$$\times \begin{cases} i^{2l}\xi^{-2l-1}\left(\frac{kr}{2B}\right)^{2l}\exp(-i2l\alpha) & \text{case a} \\ l!\xi^{-l-1}L_l^0(\frac{k^2 r^2}{4\xi B^2}) & \text{case c} \end{cases} \quad (16)$$

where $w_0$ is the beam waist of the pump, $k$ and $\lambda$ are the respective wavenumber and wavelength of the SHG beam, $L_l^0$ generalized Laguerre polynomials, and

$$\xi = \frac{2}{w_0^2} + \frac{ikA}{B}. \quad (17)$$

In the above derivations, we have used the following integral formulas [5]

$$\int_0^{2\pi}\exp[-in\theta_1 + ikbr\cos(\theta_1 - \theta_2)]d\theta_1$$
$$= 2\pi\exp[in(\frac{\pi}{2}-\theta_2)]J_n(kbr) \quad (18)$$

$$\int_0^{\infty}\exp(-ax^2)J_\nu(2bx)x^{2n+\nu+1}dx$$
$$= \frac{n!}{2}b^\nu a^{-n-\nu-1}\exp(-\frac{b^2}{a})L_n^\nu(\frac{b^2}{a}) \quad (19)$$

The ABCD transfer matrix for free space over distance $z$ reads as

$$\begin{pmatrix} A & B \\ C & D \end{pmatrix} = \begin{pmatrix} 1 & z \\ 0 & 1 \end{pmatrix}. \quad (19)$$

Inserting equation (19) into (16), and using the far-field approximation ($z \gg kw_0^2/4$), equation (16) reduces to

$$E_{SHG}(r,\alpha,z) = \frac{i}{\lambda z}\exp(-ikz)\exp(-\frac{r^2}{w^2}) \times \begin{cases} \frac{i^{2l}}{l!}\left(\frac{\sqrt{2}r}{w}\right)^{2l}\exp(-i2l\alpha) & \text{case a} \\ L_l^0(\frac{r^2}{w^2}) & \text{case c} \end{cases} \quad (20)$$

$$\propto \begin{cases} LG_0^{2l}(r,\alpha,z) & \text{case a} \\ LG_l^0(r,\alpha,z) & \text{case c} \end{cases}$$

where $w = \frac{zw_0}{\sqrt{2}z_R}$ and $z_R = \frac{kw_0^2}{4}$ are the spot radius of the SHG beam at $z$ and the Rayleigh range, respectively. The beam waist of the SHG beam is $\frac{w_0}{\sqrt{2}}$. Equation (20) shows that the generated SHG beams will evolve into modes $LG_0^{2l}$ (case a), $LG_0^{-2l}$ (case b), and $LG_l^0$ (case c) in the far-field.